\documentclass[prd,showkeys,superscriptaddress,twocolumn]{revtex4}
\usepackage{epsfig}
\usepackage{graphicx}
\usepackage{amsmath,amssymb,amsfonts,latexsym}
\usepackage{graphicx}

\usepackage{epsfig,amssymb,amsfonts,verbatim}

\usepackage{amsmath}
\usepackage{latexsym}
\usepackage{amsfonts}
\usepackage{amssymb}
\usepackage{color}

\def\bfl{\begin{flushleft}}
\def\efl{\end{flushleft}}
\def\bfr{\begin{flushright}}
\def\efr{\end{flushright}}
\def\bc{\begin{center}}
\def\ec{\end{center}}

\def\ba{\begin{eqnarray}}
\def\ea{\end{eqnarray}}
\def\baa#1{\begin{array}{#1}}
\def\eaa{\end{array}}
\def\bw{\begin{widetext}}
\def\ew{\end{widetext}}

\def\text#1{\mbox{#1}}

\begin{document}

\title{Electron-conduction mechanism and specific heat above transition temperature in LaFeAsO and BaFe$_{2}$As$_2$ superconductors}

\author{Andrew Das Arulsamy}
\email{andrew@physics.usyd.edu.au}
\affiliation{School of Physics, The University of Sydney, Sydney, New South Wales 2006, Australia}

\author{Kostya (Ken) Ostrikov}
\affiliation{CSIRO Materials Science and Engineering, P.O. Box 218, Lindfield NSW 2070, Australia}

\date{\today}

\begin{abstract}
The ionization energy theory is used to calculate the evolution of the resistivity and specific heat curves with respect to different doping elements in the recently discovered superconducting Pnictide materials. Electron-conduction mechanism in the Pnictides above the structural transition temperature is explained unambiguously, which is also consistent with other strongly correlated matter, such as cuprates, manganites, titanates and magnetic semiconductors. 
\end{abstract}


\maketitle

\section{Introduction}

The discovery of iron-based layered compound~\cite{zim,zim2,zim3} that has been found to be superconducting recently~\cite{kami} have led to detailed electrical~\cite{taka}, structural and magnetic~\cite{jun} characterizations. The superconducting transition temperatures ($T_c$) for this compound, namely, SmFeAsO$_{0.85}$F$_{0.15}$, LaFeAsO$_{1-x}$F$_x$ and CeFeAsO$_{1-x}$F$_x$ are in the vicinity of 42 K, 26 K and 30 to 43 K, respectively~\cite{chen,chen2,kami2,jun,sef,pra,pra2}. One of the fundamental requirements from the electronic characterizations of superconductors is to understand what parameters effectively control the flow of electrons, both above and below $T_c$. Similar to high-$T_c$ cuprate superconductors~\cite{bed}, the pairing mechanism in this new class of pnictide materials is also unknown presently~\cite{sa}. Studies on this class of material may provide some additional clues on the overall structure of the theory of superconductivity. When $x$ is varied via substitutional doping, for example, substituting F into O sites, two important but unexplained effects are observed, (i) systematic change in $T_c$ and (ii) systematic change in the resistivity (or resistance) above $T_c$ and above structural transition temperature ($T_N$). 

In this work, we focus on point (ii) and evaluate the mechanism of electron-conduction above $T_N$ for different substitutional doping, $x$ in (Sm, Ce, La, Ba, Sr, Ca)-FeAsO$_{1-x}$-(F, Cl, Br)$_x$ and Ba$_{1-x}$-(Sr, Ca, K, Rb, Cs)$_x$-Fe$_{2}$As$_2$ superconductors. Here, we provide unambiguous and consistent explanations on the electron-conduction mechanism based on the ionization energy theory~\cite{a1,a2}, which can be used to understand how and why both cationic and anionic substitutional doping change the resistivity and specific heat of these superconductors at temperatures above $T_N$. The advantage of this theory is that we do not need to know the microscopic details of the compound and therefore, the ionization energy theory (IET) is not suitable to evaluate the mechanism of superconductivity as stated in (i). 

\section{Ionization energy theory}

Electrons are indistinguishable, but the energy of each electron is different in different solids and if we can exploit this energy, then we can systematically study the changes of electronic properties in non free-electron metals. This is also true for electrons in atoms due to different magnitudes of electron-nucleus Coulomb force. In view of this, we propose here a methodology that interrogates the atoms to reveal information about their electronic energies based on the many-body Hamiltonian,

\begin {eqnarray}
\hat{H}\varphi = (E_0 \pm \xi)\varphi. \label{eq:1}
\end {eqnarray}

$\hat{H}$ is the many-body Hamiltonian, while $\varphi$ is the many-body wavefunction. The eigenvalue, $E_0 \pm \xi$ is the total energy, in which $E_0$ is the total energy at temperature ($T$) equals zero and $\xi$ is the atomic energy-level difference. The + sign of $\pm\xi$
is for the electron ($0 \rightarrow +\infty$) while the $-$ sign is for the hole ($-\infty \rightarrow 0$). Complete details about this theory and proofs can be found in Refs.~\cite{a1,a2} and references therein. The average ionization energy value (after invoking the ionization energy approximation) can be obtained from

\begin{eqnarray}
\xi \propto E_I = \sum_i^z \frac{E_{Ii}}{z}. \label{eq:3}
\end{eqnarray}

The subscripts, $i$ = 1, 2,...$z$, where $z$ denotes the number of valence electrons that can be excited or contributes to the electronic properties of a solid. In addition, the carrier density ($n$) can be calculated from 

\begin{eqnarray}
n = \int_0^\infty{f_e(E_0,E_I)N_e(E_0)dE_0}~\propto ~\exp{\bigg[\frac{-E_I}{k_BT}\bigg]}, \label{eq:4}
\end{eqnarray} 

where $f_e(E_0,E_I)$ is the ionization energy based Fermi-Dirac statistics, $k_B$ denotes the Boltzmann constant whereas, $N_e(E_0)$ is the density of states. The specific heat formula is given by

\begin {eqnarray}
C_v = \frac{2\pi^2k_B}{5} \bigg[\frac{k_BT}{\hbar c}\bigg]^3 e^{-\frac{3}{2}\lambda(\xi - E_F^0)} ~\propto ~\exp{\bigg[-\frac{3}{2}\lambda E_I\bigg]}, \label{eq:5}
\end {eqnarray}

where $e$ is the electronic charge, $\lambda = (12\pi\epsilon_0/e^2)a_B$, $\hbar$ is Planck constant, $c$ is the sound velocity, $\epsilon_0$ and $a_B$ denote permittivity of free space and atomic hydrogen Bohr radius, respectively. Hence, all we need to know now is the relationship between $E_I$ and $x$ in order to predict the evolution of carrier density (conductivity) and the specific heat above $T_N$. 

\section{Analysis and predictions}

From the structural studies~\cite{zim}, the oxidation (valence) states for the elements in (Sm,Ce,La)FeAsO$_{1-x}$F$_x$ are given by Sm$^{3+}$, Ce$^{3+}$, La$^{3+}$, Fe$^{2+}$, As$^{3-}$, O$^{2-}$ and F$^{-}$. Fluorine doped CeFeAsO$_{1-x}$F$_x$ is equivalent to electron doping in which, the charge distribution between the planes can be written as (FeAs)$^{-1-\delta}$ and ((La,Ce)O$_{1-x}$F$_{x}$)$^{1+\delta}$, where the $\delta$ refers to the average number of electrons generated by Ce or La due to F doping. For example, $\delta$ = 1 for $x$ = 1, and $\delta$ = 0 if $x$ = 0, where $x$ can be related to any anions with 1$-$ valence state. These additional electrons then give rise to conduction in the FeAs planes~\cite{taka,jun}. Our present understanding is that Ce or La will contribute to an additional electron in the (FeAs)$^{-1-\delta}$ plane due to F doping, which enhances the conductivity above $T_N$. However, in order to discriminate the conductivity between Ce and La (or any other 3+ valence state doping elements), we need to employ the ionization energy theory introduced earlier. In addition, this theory can be used for any substitutional elements, be it anionic as in LaFeAsO$_{1-x}$F$_x$, or cationic in both (Sm,Ce,La)FeAsO and Ba$_{1-x}$K$_{x}$Fe$_{2}$As$_{2}$~[\onlinecite{ni,ni2,ni3}].            

\begin{figure}[hbtp!]
\begin{center}
\scalebox{0.4}{\includegraphics{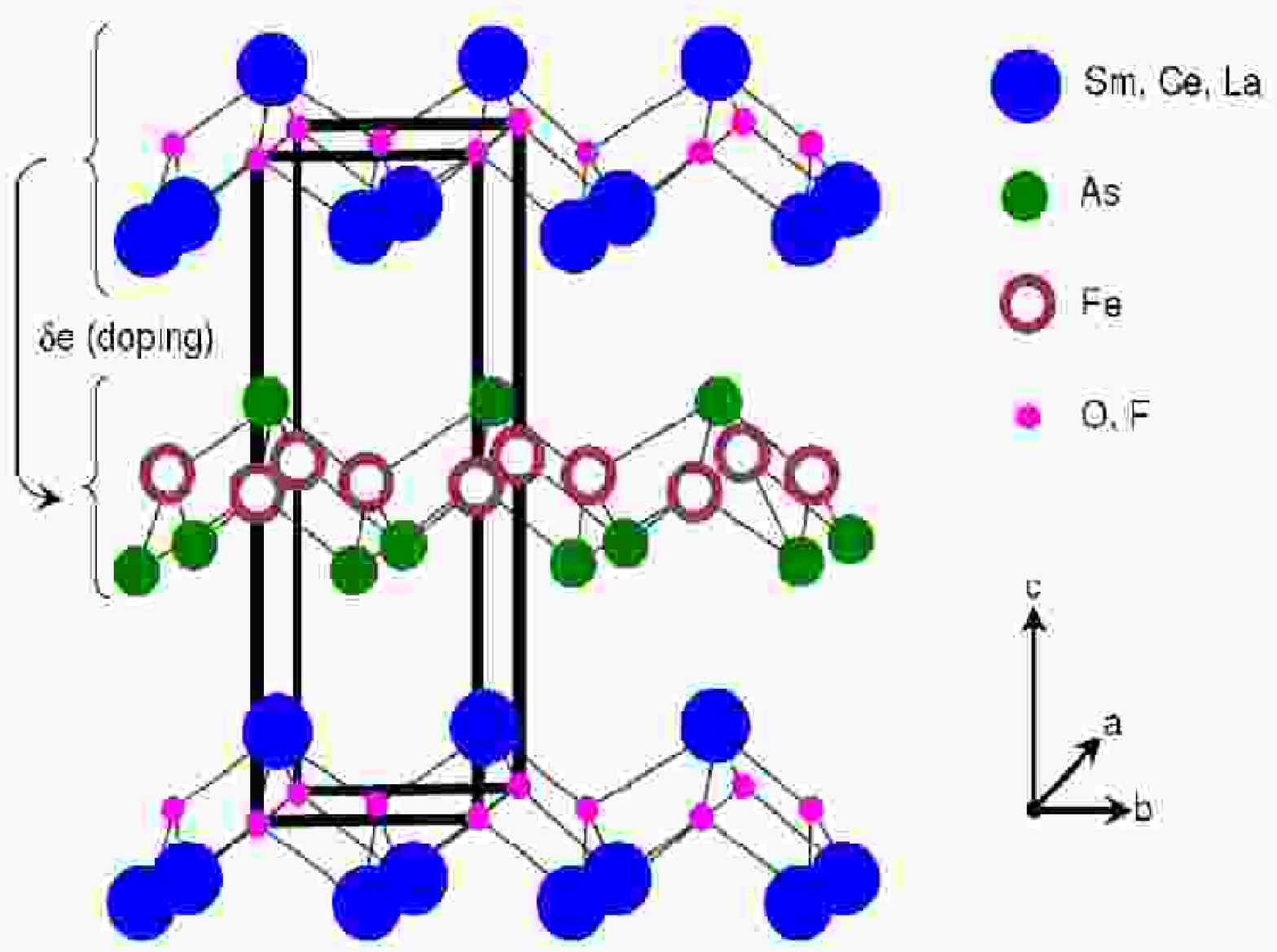}}
\caption{Schematic crystal structure for (Sm,Ce,La)FeAsO$_{1-x}$F$_x$ compound~\cite{chen2}. The source of the electrons (Sm, Ce, La) determines the ability to conduct electricity and these electrons availability comes from F doping. FeAs is the conducting layer, whereas (Sm, Ce, La)-(O,F) is the carrier doping layer. The $\delta$e doping into the FeAs layer here is due to F substitution into O sites in the layer, (Sm,Ce,La)O$_{1-x}$F$_x$.}   
\label{fig:1}
\end{center}
\end{figure}

\begin{table}[ht]
\caption{Valence states and the ionization-energy values of selected elements (ordered with respect to increasing atomic number, $Z$). The average ionization energies were calculated from Eqn.~(\ref{eq:3}). The unit kJ/mol is adopted for numerical convenience.} 
\begin{tabular}{l c c c } 
\hline\hline 
\multicolumn{1}{l}{Elements} &   ~~~~Atomic number  & ~~~~Valence & ~~~~Ionization energy \\  
\multicolumn{1}{l}{}         &    ~~~~$Z$             & ~~~~states  & ~~~~$E_I$ (kJ/mol) \\  
\hline 

F                                   &  9 						 &  1+              &  1681         \\ 
Cl                                  &  17						 &  1+              &  1251         \\ 
K                                   &  19						 &  1+              &  419          \\
Ca                                  &  20						 &  2+              &  868          \\ 
Fe                                  &  26						 &  2+              &  1162         \\ 
Br                                  &  35						 &  1+              &  1140         \\ 
Rb                                  &  37 					 &  1+              &  403          \\ 
Sr                                  &  38						 &  2+              &  807          \\
Cs                                  &  55						 &  1+              &  376          \\ 
Ba                                  &  56						 &  2+              &  734          \\ 
La                                  &  57						 &  3+              &  1152         \\ 
Ce                                  &  58						 &  3+              &  1177         \\
Sm                                  &  62						 &  3+              &  1292         \\

\hline  
\end{tabular}
\label{Table:I} 
\end{table}

\begin{figure}[hbtp!]
\begin{center}
\scalebox{0.4}{\includegraphics{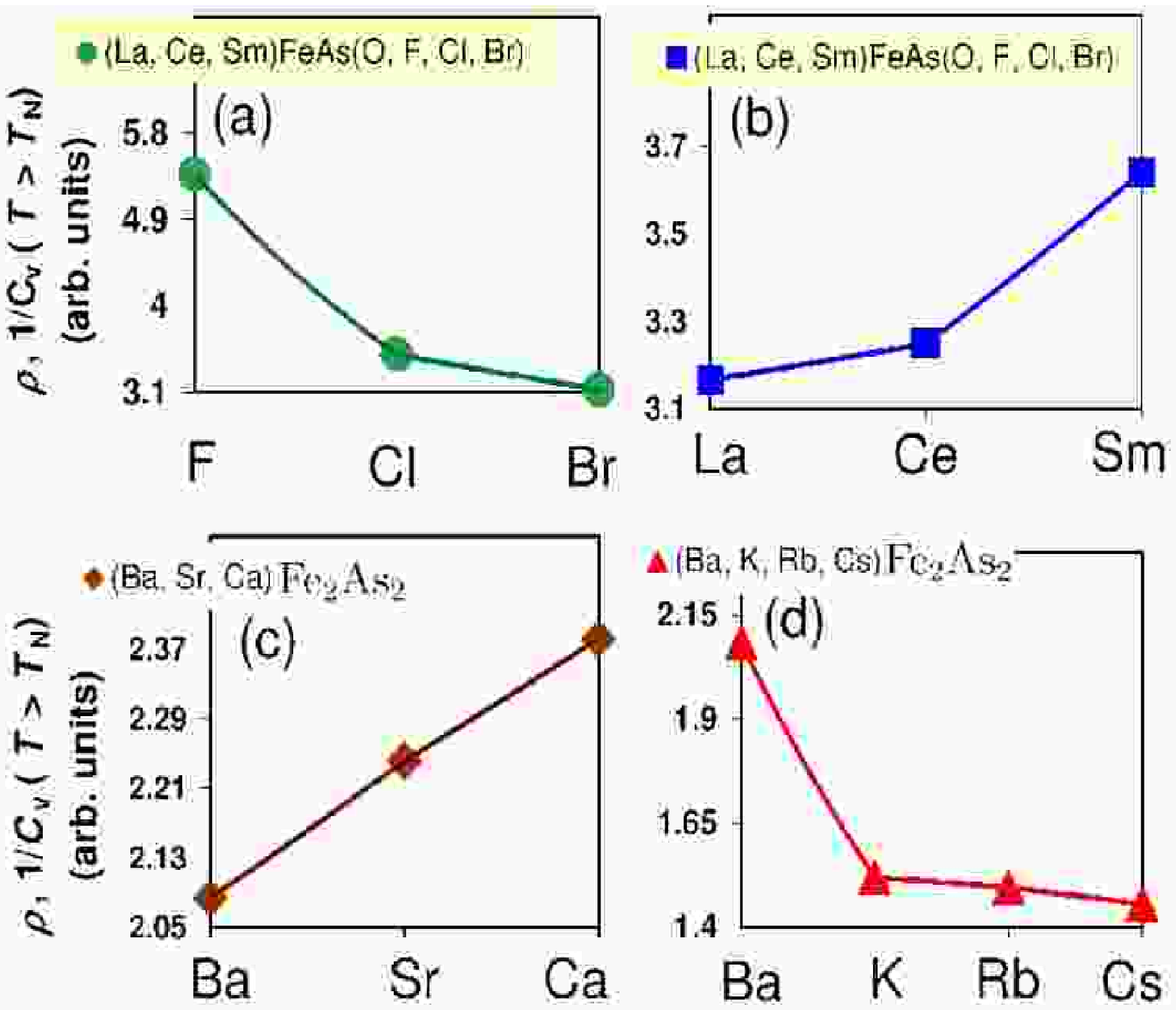}}
\caption{Predicted evolution of the resistivity ($\rho$) and inverse specific heat ($C_v$) curves (upward or downward) with respect to different doping elements (both cations and anions) based on the ionization energy theory. The predicted shifts depicted in (a) to (d) are valid for temperatures, $T > T_N$. The predicted trends for resistivity and specific heat were obtained from Eqn.~(\ref{eq:4}) and Eqn.~(\ref{eq:5}). Equation~(\ref{eq:3}) is used to average the ionization energies of all the elements. See text for details.}   
\label{fig:2}
\end{center}
\end{figure}

Let us start examining the conduction mechanism in (Sm,Ce,La)FeAsO$_{1-x}$F$_x$ compound (see Fig.~\ref{fig:1}) due to increasing F (and decreasing O). Using Eqn.~(\ref{eq:3}), we have calculated the averaged ionization-energy values for all the elements discussed here are listed in Table~\ref{Table:I}. Prior to averaging, the ionization energies for all the elements were obtained from Ref.~\cite{web28}. We stress here that smaller $E_I$ implies weak electron-phonon coupling that gives rise to easier electron-flow and also large carrier density~\cite{a1}. However, this scenario is reversed if $E_I$ is large. This means that systematic substitution of elements with smaller $E_I$ will systematically decrease the $E_I$ of the compound and subsequently increases the carrier density due to increased electronic excitation probability. In principle, we can use this theory to fine-tune the conductivity of any strongly correlated matter. Firstly, we need to identify the source of the electrons, which is from the carrier doping layer, LaO$_{1-x}$F$_x$. Since the valence state for F is 1$-$, and on average, this increases the electrons from LaO$_{1-x}$F$_x$ layer into the (FeAs)$^{-1-\delta}$ plane with respect to increasing F~[\onlinecite{taka,jun}]. The next question is how does substitutional doping of Cl or Br, instead of F will affect the resistivity and specific heat curves? Similarly, Sm or Ce substitution into La sites will also affect these curves systematically that needs to be understood. To answer these questions unequivocally, we need to invoke the ionization energy theory, and its predictions are given in Fig.~\ref{fig:2}. 

We first discuss the electron conduction mechanism and the evolution of specific heat in the (Sm,Ce,La)FeAsO$_{1-x}$F$_x$ and Ba$_{1-x}$K$_{x}$Fe$_{2}$As$_{2}$ compounds, for different doping elements. Since F, Cl and Br are all anions, we are only required to know the first ionization energies of these atoms to evaluate the upward and/or downward shifts of the resistivity curve~\cite{a1}. The reason that we used the first ionization energy is that the atoms with large $E_I$ (F in this case) implies strong Coulomb attraction (due to smaller screening) between an electron and the ions. From Table~\ref{Table:I}, we can write $E_I$(F) $>$ $E_I$(Cl) $>$ $E_I$(Br), and after substituting this inequality into the ionization energy based Fermi-Dirac statistics~\cite{a1} and using Eqn.~(\ref{eq:4}), we can obtain the results where the resistivity curve will shift downward (improved conductivity above $T_N$) with increasing content of Cl or Br (see Fig.~\ref{fig:2}(a)). The reason is that when the elements with smaller ionization energies are substituted, then the electronic excitation probability will increase, which subsequently leads to increased carrier density or improved conductivity, and larger specific heat. Therefore, the resistivity curve ($\rho(T > T_N)$) shifts downward, while the specific heat curve shifts upward systematically with Cl or Br doping as in O$_{1-x}$Cl$_{x}$ or O$_{1-x}$Br$_{x}$, as compared to O$_{1-x}$F$_{x}$ (see Figs.~\ref{fig:2}(a) and (b)). In other words, Br-doped samples will have the lowest resistivity, whereas F-doped sample will have the highest resistivity. The reason we used $T > T_N$ is because when $T < T_N$, the electronic excitation or the electronic polarization probability of the compound changes. Consequently, any change in these probabilities also corresponds to the change in the average valence states of all the elements in the compound. This is the reason why we see the structural changes (changes in the lattice parameters). 

\begin{figure}[hbtp!]
\begin{center}
\scalebox{0.4}{\includegraphics{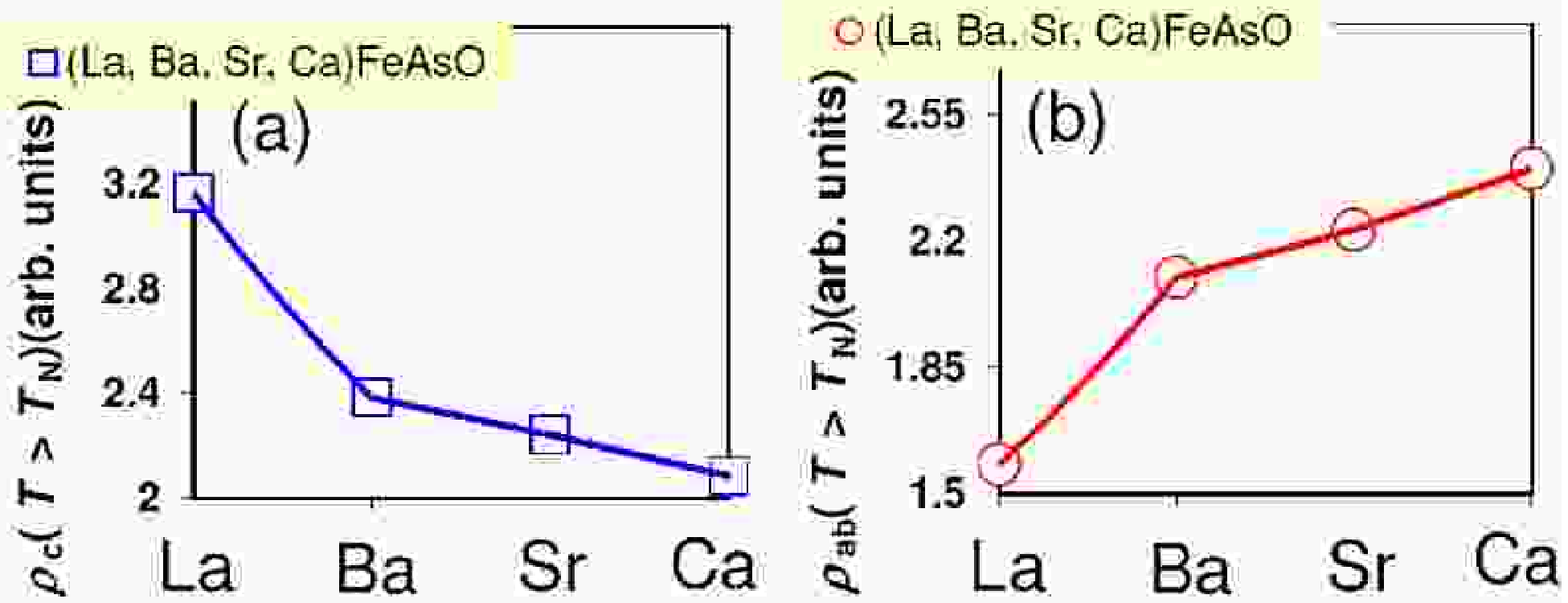}}
\caption{Theoretically calculated resistivities in (a) $c-$axis (shifts downward) and (b) $ab-$plane (shifts upward) for La$_{1-x}$(Ca,Sr,Ba)$_x$FeAsO compound. Note that these predictions are observable if $\rho_{\rm{layer}}^{\rm{LaO}} > \rho_{\rm{layer}}^{\rm{FeAs}}$, otherwise, $\rho_c \propto \rho_{ab}$.}   
\label{fig:3}
\end{center}
\end{figure}

On the other hand, if we were to substitute La with Ce or Sm, the resistivity and specific heat curves will shift upward and downward, respectively. In all the plots depicted in Fig.~\ref{fig:2}, the $y-$axis denotes $\rho(T > T_N)$, $1/C_v$ where $\rho \propto 1/n \propto 1/C_v$ based on Eqns.~(\ref{eq:4}) and~(\ref{eq:5}). It is very important to realize here that there are different sets of Eqns.~(\ref{eq:4}) and~(\ref{eq:5}), of which, one set of equations captures the changes from F to Cl doping in Fig.~\ref{fig:2}(a) and another set of equations are required to capture the changes from Cl to Br. The reason for this is that the electronic and phononic properties of a given sample in the presence of O and F has its own exponential trend that are different from the samples with O and Cl, or O and Br, and so on for other elements shown in Figs.~\ref{fig:2}(b)$-$(d). Add to that, in real sample preparations for different doping levels ($x$), the valence states of all the elements may not remain the same for different $x$ due to defects (interstitial, substitution at different sites and vacancies)~\cite{a2}, which further affect the exponential term given in Eqns.~(\ref{eq:4}) and~(\ref{eq:5}). As a consequence, all the solid lines connecting each calculated points in Figs.~\ref{fig:2} and~\ref{fig:3} have their own exponential terms based on Eqns.~(\ref{eq:4}) and~(\ref{eq:5}), and therefore they are discontinuous.

Using the same line of reasoning based on the ionization energy theory discussed above, we can also explain and predict the effect of substitutional doping in Ba$_{1-x}$(Sr,Ca,K,Rb,Cs)$_{x}$Fe$_{2}$As$_{2}$ compound. The upward shift in the resistivity and downward shift for the specific heat curves are clearly shown in Fig.~\ref{fig:2}(c) when Ba is substitutionally doped with Sr or Ca. Moreover, Fig.~\ref{fig:2}(d) also indicates the effect of cation-doping at Ba sites, but these cations have 1+ valence state in which, K, Rb or Cs will give rise to improved conductivity above $T_N$, and larger heat capacity, as compared to Ba. 

We stress here that the theoretical results presented in Fig.~\ref{fig:2} are valid for both $ab-$plane and $c-$axis resistivities, which implies that they are observable even for polycrystalline samples. However, the analysis will become complicated for the La$_{1-x}$(Ca,Sr,Ba)$_x$FeAsO compound due to different conductivities in FeAs and LaO layers. In contrast, any substitution into Ba sites in BaFe$_{2}$As$_{2}$ compound will affect both the $ab-$plane and $c-$axis resistivities equally (upward or downward: $\rho_{ab} \propto \rho_c$) because there are no anions surrounding the Ba. Unlike BaFe$_{2}$As$_{2}$ compound, the LaFeAsO compound can be substitutionally doped in such a way to give contradicting resistivity curve shifts or trends between $ab-$plane and $c-$axis resistivities. These contradicting trends can be observed (i) if La is substitutionally doped with elements that have average valence states lower than 3+, (ii) elements with lower average ionization energies than La$^{3+}$ are substituted and (iii) $\rho_{\rm{layer}}^{\rm{LaO}} > \rho_{\rm{layer}}^{\rm{FeAs}}$. If condition (iii) is not satisfied, then $\rho_c \propto \rho_{ab}$, where both resistivities will have the same upward or downward shifts. For example, substituting La$^{3+}$ with Ba$^{2+}$ or Sr$^{2+}$ or Ca$^{2+}$ in LaFeAsO compound will improve the electron conduction in the La$_{1-x}$(Ba,Sr,Ca)$_x$O layers, while the resistance in the FeAs layers will increase due to the lack of electron-doping from the La$_{1-x}$(Ba,Sr,Ca)$_x$O layers. Hence, by measuring the resistivities in $ab-$plane and $c-$axis of a single crystal, we will see a downward shift in the $c-$axis resistivity due to improved conductivity in La$_{1-x}$(Ba,Sr,Ca)$_x$O layers. Whereas, an upward shift is expected for $ab-$plane resistivities due to smaller electron doping from La$_{1-x}$(Ba,Sr,Ca)$_x$O layers. The predicted results based on the ionization energy theory are shown in Figs.~\ref{fig:3}(a) and (b).

Thus far, there are two experimental results found to be in accordance with our predictions. Firstly, the resistivity curves of the polycrystalline Ba$_{1-x}$K$_{x}$Fe$_{2}$As$_{2}$ samples indeed have a downward shift with doping $x$~[\onlinecite{ni2}], as correctly predicted in Fig.~\ref{fig:2}(d) (where K doped samples have a lower resistivity at $T > T_N$). Apart from that, the polycrystalline Sr doped La$_{1-x}$Sr$_x$FeAsO samples shows an upward systematic-shift in the resistivity curve from $x$ = 0.13 to 0.20~[\onlinecite{wen}]. This latter results indicate that La$_{1-x}$Sr$_x$FeAsO samples satisfy condition (iii), where $\rho_{\rm{layer}}^{\rm{LaO}} > \rho_{\rm{layer}}^{\rm{FeAs}}$. Based on the principle of least action, the resistivity measurements in polycrystalline samples will be determined by the lowest resistance path~\cite{a1}, which corresponds to $\rho_{\rm{layer}}^{\rm{FeAs}}$ in this case. Therefore, Sr doped La$_{1-x}$Sr$_x$FeAsO single crystals are predicted to obey the theoretical results displayed in Figs.~\ref{fig:3}(a) and (b).      

\section{Conclusions}

In conclusion, we have provided a physical picture based on the ionization energy to predict and explain the electron conduction in (Sm, Ce, La, Ba, Sr, Ca)-FeAsO$_{1-x}$-(F, Cl, Br)$_x$ and Ba$_{1-x}$-(K, Rb, Cs)$_x$-Fe$_{2}$As$_2$ superconductors above the structural transition temperature. Our predictions can be tested with the present experimental techniques. Accurate predictions on the resistivity and specific heat has been made by identifying the type of ions (to obtain their respective energy-level difference) and their valence states (to transform the atomic energy-level difference to the ionization energy) in the compound. By identifying the type of ions that exist in a given compound, we can evaluate the conductivity and the specific heat of the compounds accurately by means of the ionization energies. However, IET, in its present form, cannot be used to understand the origin of superconductivity. 

\section*{Acknowledgments}

A.D.A. would like to thank the School of Physics, University of Sydney for the USIRS award, and Kithriammah Soosay for the partial financial support. K.O. acknowledges the partial support from the Australian Research Council (ARC) and the CSIRO.

\end{document}